\documentclass[11pt,twoside]{article}

\usepackage{asp2006}
\usepackage{epsf}
\usepackage{psfig}
\usepackage{lscape}

\begin{document}
\title{Dynamics of the isolated galaxy CIG\,0314}   
\author{S. Verley$^{1,2,3}$, M. Rosado$^2$, P. Repetto$^2$, R. Gabbasov$^2$, L. Verdes-Montenegro$^3$, G. Bergond$^4$, I. Fuentes-Carrera$^5$, A. Durbala$^6$}   
\affil{$^1$Universidad de Granada, $^2$IA-UNAM, $^3$IAA, $^4$CAHA, $^5$IPN, $^6$UA}    

\begin{abstract} 
In the context of the AMIGA project, we used Fabry-Perot observations in order to study the dynamics of the ionised gas in the isolated galaxy CIG\,0314. From the H$\alpha$ observations, we could obtain the velocity field and rotation curve of the galaxy. A detail analysis of the velocity field is done in order to understand the kinematics of the gas to gather clues on the mechanisms which favour or inhibit star formation, in particular along the bar. The visible and dark matter content can be reached, as well as an estimation of the mass of the galaxy.
\end{abstract}


In the framework of the AMIGA project (Verdes-Montenegro et al. 2005), we are interested in studying the dynamics of galaxies which are not perturbed by external influences. The late type CIG\,0314 (NGC\,2776) is a well isolated galaxy being in a very low density field of galaxies ($\eta_k$ = 0.984) and affected by very low values of external tidal forces (Q = -2.682, Q$_{\rm Kar}$ = -3.719), as was verified in Verley et al. (2007a,b). H$\alpha$ observations of CIG\,0314 were made at the San Pedro Martir telescope (OAN, Mexico) making use of the scanning Fabry-Perot interferometer PUMA (Rosado et al. 1995). We used the ADHOC (Boulesteix 1993) and CIGALE (Le~Coarer et al. 1993) softwares in order to reduce and analyse the cube of H$\alpha$ data. The velocity fields and the rotation curves were derived from the Fabry-Perot observations. An estimation of the mass of the galaxy can be obtained from the rotation curves (Lequeux 1983): RV$^2$(R)/G. Assuming a distance of 37.6 Mpc, we have estimated a dynamical mass of 3.34~$\times$~10$^9$~M$_\odot$ within a radius of 4~kpc.

Besides the particular case of the dynamical study of CIG\,0314, our final aim is to study the kinematics of the ionised gas in order to understand the kinematical behaviour of the gas and its relation with star formation in isolated galaxies (CIG/AMIGA catalogue). This will be in order to interpret the evolutive sequence proposed in Verley et al. (2007c).


\acknowledgements 
We thank the organisers of the conference.


\end{document}